\documentclass[10pt,journal,letterpaper]{IEEEtran}
\usepackage[binary]{SIunits}
\usepackage[cmex10]{amsmath}
\usepackage[dvips]{graphicx}
\usepackage{color}
\usepackage{cite}
\usepackage{url}
\usepackage[caption=false]{subfig}
\usepackage{multirow}

\renewenvironment{IEEEbiography}[1]
  {\IEEEbiographynophoto{#1}}
  {\endIEEEbiographynophoto}

\title{LTE enhancements for Public Safety and Security communications to support Group Multimedia Communications}

\author{Lorenzo~Carl\`a, Romano~Fantacci,~\IEEEmembership{Fellow,~IEEE,} Francesco~Gei, Dania~Marabissi,~\IEEEmembership{Senior Member,~IEEE,} and~Luigia~Micciullo
 \thanks{This work was supported by Wilife project}\\
 \thanks{All the authors are with the Department of Information Engineering (University of Florence, Italy).}}

\begin{document}
\maketitle
\begin{abstract}
Currently Public Safety and Security communication systems rely on reliable and secure Professional Mobile Radio (PMR) Networks that are mainly devoted to provide voice services. However, the evolution trend for PMR networks is towards the provision of new value-added multimedia services such as video streaming, in order to improve the situational awareness and enhance the life-saving operations. The challenge here is to exploit the future commercial broadband networks to deliver voice and multimedia services satisfying the PMR service requirements. In particular, a viable solution till now seems that of adapting the new Long Term Evolution technology to provide IP-based broadband services with the security and reliability typical of PMR networks. This paper outlines different alternatives to achieve this goal and, in particular, proposes a proper solution for providing multimedia services with PMR standards over commercial LTE networks.
\end{abstract}


\section{Introduction}
Worldwide there is a great interest of governments and organizations involved in Public Safety and Security (PSS) towards the evolution of existing wireless systems for critical communications based on Professional Mobile Radio (PMR) technologies such as Terrestrial Trunked Radio (TETRA), TETRA for Police (TETRAPOL), or Association of Public-Safety Communications Officials-Project 25 (APCO P25).
It is widely recognized that efficient communications are of paramount importance to deal with emergency or disaster situations. The possibility of PSS operators to use a wide range of data-centric services, such as video sharing, files transmission (e.g., maps, databases, pictures), ubiquitous
internet and intranet access, has a strong impact on the efficiency and the responsiveness of the emergency services.
However, PMR systems are based currently on 2G networks, thus offering a wide range of voice services, but having a limited possibility to support data services. 
Even if some efforts have been done to enhance PMR systems and to offer higher communication capacity, achievements are still behind those made in the commercial world that recently has developed the 3GPP Long Term Evolution (LTE) technology. 
Hence, there is a great consensus in adopting the commercial LTE framework to answer to the PSS communication needs. 
A common standard for commercial and PSS environments can open the door to new opportunities and synergies,
offering advantages to both worlds. 
However, the LTE technology needs some specific enhancements to be fully compliant with the mission and  safety critical requirements. Indeed, PMR networks must be reliable, secure and resilient, guaranteeing service accessibility and wide coverage as well. In addition PSS operators need some specific applications and functionalities such as push-to-talk, dispatch services, priority management, group communications and direct communications~\cite{tetra_book}. Therefore, adopting LTE as PMR broadband technology needs that these features are included in the future releases of the 3GPP standard also guaranteeing interoperability with actual narrowband PMR systems.  
In this paper, Sec.~\ref{REQ} analyses the main requirements of PMR communications. The use of LTE for critical communications and the description of critical services currently not supported by the LTE standard is given in Sec.~\ref{LTE-PSS} underling the need of further research activity. Sec.~\ref{AP} proposes an evolution path of the PMR LTE-based network architecture. Finally the attention is devoted to group communications in Sec.~\ref{GC}, where the recommendations that are under evaluation by the 3GPP are analysed and on their basis our proposal is detailed. Conclusions are drawn in Sec.~\ref{CL}.

\section{PSS communication system requirements}\label{REQ}

Mission critical communications are characterized by different and more severe requirements respect to commercial communications. In particular, we can distinguish between typical narrowband PMR and new broadband requirements and functionalities:
\begin{itemize}
\item \textbf{PMR narrowband requirements and functionalities} \cite{tetra_book}
\begin{itemize}
\item \textit{High reliability and availability.} The system shall be available for 99\% of time in 24 hours and 99.9\% of time in a year. It shall cover 96\% of the area in outdoor environment and 65\% in indoor. 
\item \textit{Half-duplex and full-duplex voice calls}.
\item \textit{Fast call setup time} lower than 300 $\milli\second$ \cite{22468}.
\item \textit{Push to Talk} management in half-duplex communications.
\item \textit{Call priority and preemption}. The system shall assign different levels of priority to calls and interrupt low priority calls on arrival of high priority calls that do not find available resources.
\item \textit{Direct Mode Communications} between terminals without the support of network infrastructure.
\item  \textit{Text message service}, e.g., the Short Data Service of TETRA system.
\item \textit{Network interoperability}. Communications with users located on \textit{external networks} or PSTN (Public Switched Telephone Network).
\item \textit{Emergency calls}.
\item \textit{Group calls}.  GCs allow a user to talk simultaneously with several users belonging to a group that can be 
predefined or formed on-demand.
 The network shall be able to permit the coexistence of many active groups at the same time. 
A real life scenario comprises an average of 36 voice groups corresponding at least to 2000 users in an area. It is expected that up to 500 users can participate in a group \cite{22468}. Moreover, each user can be registered to many groups at same time.
\end{itemize}
\item \textbf{PMR broadband requirements and functionalities} \cite{lewpMatrix}
\begin{itemize}
\item \textit{Data communications} for fax and image transfer. 
\item  \textit{Synchronous video transmission} that consists of a bidirectional communication composed of different data, audio and signaling flows at 256 kbps. 
\end{itemize} 
\end{itemize}
In order to fulfil all the above requirements and functionalities it has been evaluated that a bandwidth of 10MHz is needed \cite{lewpMatrix}. As a consequence one open point is to identify an harmonised frequency band allocation for the new PMR systems that takes into account different national allocation policies, interferences towards other existing systems and economic convenience.

\section{LTE for PSS communications}\label{LTE-PSS}
This section describes the main communication features that shall be gradually introduced in the future LTE standard releases in order to satisfy the typical needs of Public Safety organization. 
Indeed, currently the LTE system does not provide services considered vital for the PSS context. 
\subsection{Direct Communications}
Direct connection among devices (Direct Mode Operation - DMO) is a mandatory feature for PMR systems. It allows PSS users to communicate without the involvement of the network infrastructure, e.g., if the network is not available due to a failure or lack of coverage. Current PMR systems foresee the use of DMO for direct connection between user terminals (back-to-back), but also between the users and special terminals that operate as Relay and/or Gateway towards the Trunked MO (TMO) network. 
The 3GPP is working to introduce in the LTE standard the direct communication as a new service named \textit{Proximity Service} (ProSe). In particular the Technical Report \cite{TR23.703} deals with the definition of  an advanced network architecture able to support ProSe service either for commercial or PSS applications.
The service definition is on going and several proposals are under investigation. In general we can state that the ProSe foresees two different operative modes:
\begin{itemize}
\item Network Assisted
\item Not Network Assisted
\end{itemize}
In the first case the network assistance is required to authenticate terminals, allocate resources and manage the security. The network provides to terminals the set of parameters needed for the call management. In the second case, the connection is activated directly by the terminals without any network involvement, using parameters already known by the User Equipment (UE) and pre-allocated resources.
For commercial applications only the Network Assisted mode is considered, while for PSS applications both modes are allowed because in this case the ProSe shall guarantee UEs in proximity to communicate under any network condition.
In Network Assisted solutions we can also distinguish between:
\begin{itemize}
\item \textit{full network control}: the direct link among the UEs is handled by the network including control (connection set-up, maintenance) and data planes. The communications occur on licensed bands and the resources can be allocated dynamically or in a semi-static mode by the network;
\item \textit{partial network control}: the network is involved only in the authentication and authorization phase, but the direct connections between the UEs are initialized in autonomous way. Usually the resources are pre-allocated to this kind of service. 
\end{itemize}

\begin{figure*}[t]
 \centering
\includegraphics[width=\textwidth]{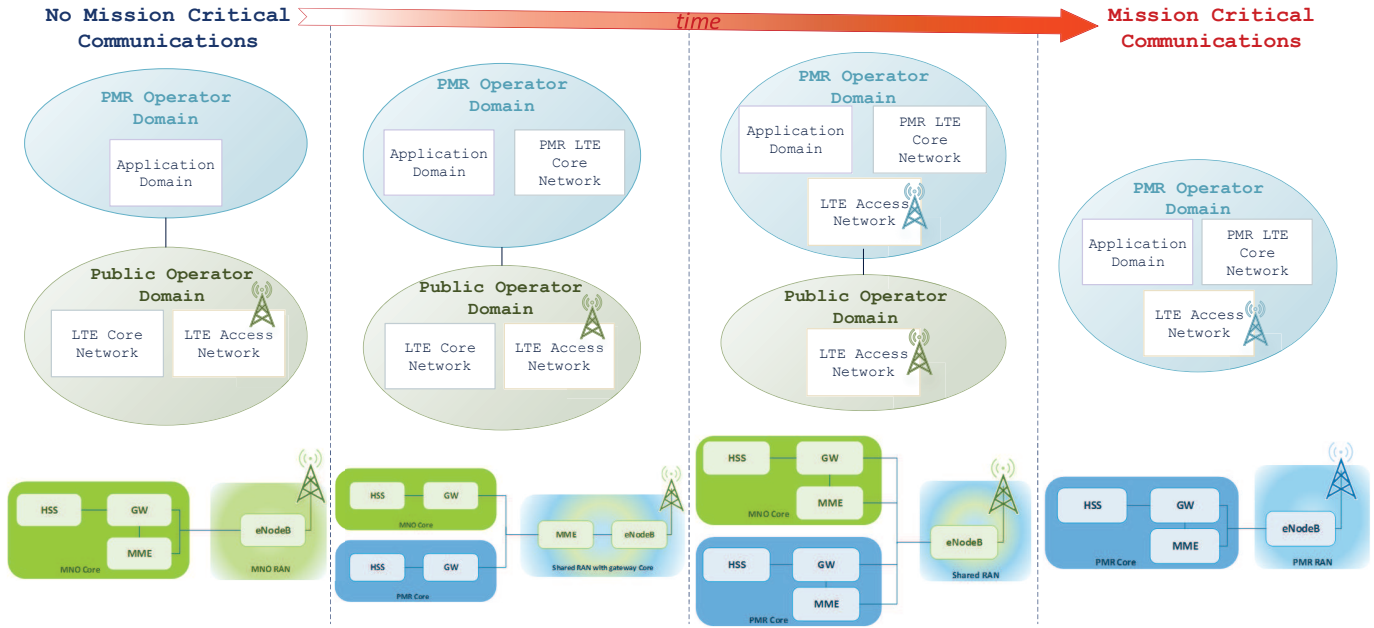}
 \caption{Evolution of a PMR network from full network sharing with commercial operator to independent professional network}
 \label{network_evol}
\end{figure*} 

\subsection{Voice service}
Even if broadband services are gaining importance in emergency and rescue operations, voice call still remain a fundamental service even for future PMR systems. However, this appears to be in conflict with the LTE architecture that is full-IP based and mainly developed for data communications, hence, unable to support the circuit switching mode typical for classical voice calls.

In LTE the implementation of voice service is following an evolution path starting from Circuit Switching Fall Back approach that relies on legacy networks \cite{23272}. In PMR environment this implies a transition phase during which voice service will be supported by the existing PMR systems.

At a later stage, voice calls will rely on Voice over LTE (VoLTE), an advanced profile defined by Global System for Mobile Communications Association (GSMA), based on IP Multimedia Subsystem architecture (IMS). VoLTE supplies basic functions as call establishment and termination using Session Initiation Protocol (SIP), call forwarding, calling ID presentation and restriction, call-waiting and multiparty conference. However, VoLTE does not support primary functions of mission critical communications, i.e., group and direct communications. Although IMS supports Open Mobile Alliance (OMA) Push-to-Talk over Cellular (PoC), which allows up to 1000 multicast groups and 200 users in one group, it is not suitable for critical communications because of the high call set-up time that reaches 4000 $\milli\second$ in case of pre-established session and 7000 $\milli\second$ if the session has to be created \cite{23979}. Thus the adoption of PoC needs some specific network solutions\cite{23768}.
Another important aspect of VoLTE is the lack of a suitable security level due to absence of end-to-end encryption mechanism typical of PMR communications~\cite{6766087}.

Communication security is of paramount importance in PSS networks whose main aspects are encryption, authentication, provisioning and user management. However, the details concerning security are out of the scope of this paper. The interested reader can refer to~\cite{6766086,6766089,6845055} for some examples of promising approaches on how to face the main security issues in a Public safety network.

\subsection{Group Calls}
The Group Call (GC) represents one of the most important and indispensable service of PMR networks. It enables an efficient management of the rescue teams and permits to send commands to all the PSS operators in a disaster area and share information. For the above reasons, the remaining part of this paper will be devoted on the discussion and proposal of approaches on how to manage GCs in a public safety LTE-based network. 

The communications addressed to more than one user can be distributed with a Point-to-Point (P2P) approach, using one \textit{unicast} transmission for each involved user, or with Point-to-Multipoint (P2M) flows. \\
In LTE the distribution of \textit{multicast} communications is demanded to the evolved Multimedia Broadcast Multicast Service (eMBMS), introduced in the 3GPP standard starting from release 10\cite{mbms}.
In particular, it t has been mainly developed for multicast and broadcast distribution of multimedia data, such as video streaming provided by external sources that act as service providers. For this reason only a downlink channel is considered. 

\section{A Public Safety LTE Network Architecture }\label{AP}
Providing mission critical services in a public safety LTE network involves a proper network architecture solution in order to achieve and maintain the required performance and reliability levels. Hence, it easy to foresee that several research efforts and further releases of the LTE standard will be needed before this  task will be fully accomplished.
 As a consequence, in the short term the simplest way for enabling high data rate services for PMR networks based on LTE, is the use of already deployed commercial networks. 3GPP provides recommendations and technical specifications for the sharing of network devices and modules between operators, commonly used in commercial networks.

LTE network is composed by several modules, which can be collected in different logical domains:
\begin{itemize}
\item the \textit{services domain}, managing the contents and the services to be provided to UEs, injecting the traffic in the network through a service delivery platform;
\item the \textit{Evolved Packet Core} (EPC) network, mainly responsible of the control functions;
\item the \textit{Radio Access Network} (RAN), composed by the eNodeBs;
\item the \textit{Users Equipment} domain.
\end{itemize}
While the service domain shall be dedicated to PMR communications, due the specific services to be provided, the core network and the radio access network can be fully or partially shared between the professional and the public networks, as shown in Fig. \ref{network_evol}.
In an initial phase the PMR network can rely completely on a commercial network, limiting to provide non mission critical professional services. This is mainly due to the fact that the Public safety agencies cannot control the fulfilment of the quality of service requirements, as well as the reliability and availability of the network, that are fully managed by Carrier Operators \cite{TCCA_roadmap}.

From the network architectural point of view, in the first two configurations represented in the figure, the PMR operator acts as a Mobile Virtual Network Operator (MVNO), in Core Network Sharing or Full RAN Sharing configuration, respectively. The two networks partially sharing their EPC can follow the Gateway Core Network (GWCN) specifications of 3GPP \cite{23251}.
In the third configuration the professional network expands to the eUTRAN, with a partial coverage of the territory. The deep diffusion of the network on the area of interest is one of the major obstacle on the upgrade of current PMR networks over new and more efficient communication technologies. LTE allows the sharing of part of the eUTRAN between operators, following the Multiple Operator Core Network (MOCN) 3GPP specification.
The last presented configuration represents an autonomous LTE PMR network, with full radio coverage of the territory. In this case the PMR operator is the owner of the eUTRAN, but some forms of passive sharing, such as mast sharing or site sharing, can be anyway implemented for reducing costs and environment impacts of the new network deployment.

\section{Voice and Multimedia Group Calls in 3GPP} \label{GC}
As stated before future LTE-based PMR networks are expected to support bandwidth consuming applications such as video, imaging and data communications. In addition, multiple GCs 
shall be able to handle a great number of UEs at the same time and in geographical areas served by several neighbour eNBs (i.e., contiguous cells). 
In this section we analyse recommendations provided by 3GPP \cite{22468}, \cite{23768} for the deployment of the GC service in LTE networks. These focus mainly on multicast transmission mode, in particular on the LTE eMBMS framework, even if unicast transmission is taken also into account.

\subsection{Unicast vs Multicast}
The advantage of the unicast transmission scheme for the delivery of a GC service is that each unicast link can be tailored to the propagation conditions experienced by the served UE through the channel state information feedback and/or by acknowledge (ACK) messaging. 
Conversely, in the multicast transmission the most adverse propagation conditions must be always considered in order to meet specific service requirements. 
However, 
the use of the unicast mode is not recommended in scenarios involving the delivery of multimedia contents to large groups of users, as the ones expected for future PMR networks. In this case, the use of the multicast mode provides a more efficient and reliable solution. 
Indeed, the main benefit of using P2M flows is that the same content can be received by many users at the same time with a bandwidth and power consumption not dependent on the number of simultaneous users.\\

\subsection{eMBMS framework in commercial LTE networks}
The eMBMS service combines the advantages of P2M transmissions with geographical and temporal flexibility. Indeed, the multicast distribution over eMBMS is not extended necessarily on the entire network, but its scope can be limited to a small geographical area, such as a city center or a stadium, as well as large areas or regions. \\
In particular, the eMBMS framework provides two kinds of transmission schemes, namely Single Cell (SC) and Single Frequency Network (SFN). In SC-eMBMS mode, each eNB delivers the MBMS data flow independently from the others. On the other hand, in SFN-eMBMS mode multiple eNBs are synchronized in order to transmit the same physical signal at the same time. The eNBs involved in the distribution of the same multicast/broadcast flow form a \textit{MBMS area}. One eNB can belong to more than one MBMS area simultaneously. 

The SFN-eMBMS transmission mode leads to a significant improvement in terms of cell coverage and spectral efficiency, since that at the UE receiving side the signals coming from multiple eNBs are combined resulting in a higher useful power. 

\begin{figure}[tbd] 
\begin{center}
\includegraphics[width=\columnwidth]{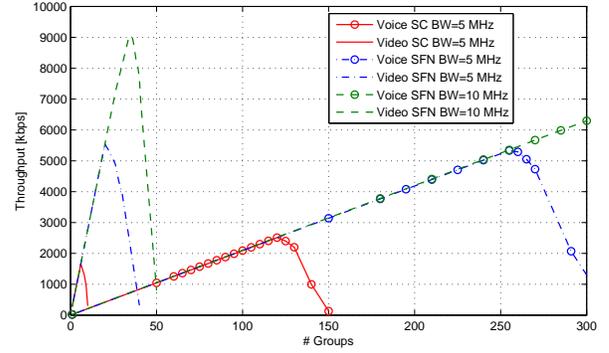}
\caption{Throughput for different types of services in SC and SFN eMBMS.}
\label{fig.rate_gc}
\end{center}
\vspace{-2mm}
\end{figure}

Fig. \ref{fig.rate_gc} shows a performance comparison between SC and SFN (with 4 eNBs) eMBMS transmission schemes in terms of system throughput as a function of the number of active Group Calls. Performance has been derived for two types of service (video flow \url{@} $256$ kbps at the application layer and voice call \url{@} $16$ kbps \cite{lewpMatrix}) and bandwidths. We can see that the throughput increases with the number of active GCs up to a maximum value. Then it decreases rapidly due to a saturation of the system. The benefits of using SFN are evident.

Another advantage of the SFN-eMBMS scheme for PMR networks is that the MBMS area can be adapted in order to cover the whole critical area involved in PSS operations.

The LTE standard schedules unicast and multicast/broadcast services in separated subframes. In particular, in each LTE radio frame up to 6 out of 10 subframes can be reserved for the delivery of eMBMS services. 

Each eMBMS service is mapped into a MBMS bearer service and into a logical \textit{Multicast Traffic Channel} (MTCH). Then the MTCHs are multiplexed with a logical Multicast Control Channel (MCCH) and mapped into a physical Multicast Channel (MCH). 

From an architectural point of view, the eMBMS framework introduces new modules for the set-up and management of multicast and broadcast contents. In particular, the \textit{Broadcast-Multicast Service Center} (BM-SC) represents the interface between the Core Network (EPC) and the \textit{Content Provider}, and it is responsible for the scheduling of MBMS services over the LTE network. The \textit{MBMS Gateway} (MBMS-GW) is a logical node that transmits the multicast flow towards the eNBs that belongs to the MBMS area. Finally, the \textit{Multicell Coordination Entity} (MCE) is a logical node that ensures that all the eNBs of the MBMS area use the same radio resource configuration.

\subsection{eMBMS support for GCs in PMR networks}
The eMBMS framework in the LTE standard is designed only for downlink distribution of multimedia data flows. Hence, it has to be modified in order to support different services such as the GC. 
To this goal 3GPP proposes to introduce a new logical interface, called \textit{Group Communications System Enabler} (GCSE), which is in charge of the creation, management and deletion of GCs. 
The proposals under evaluation~\cite{23768} are mainly based on the multicast communication approach, the unicast mode is enabled whenever a multicast transmission cannot be accomplished. In both cases all the Group Members belonging to a GC perform uplink communications through the establishment of unicast bearers.

The GC management by the GCSE must ensure also the meeting of the PMR network requirements such as 
short call latencies, service continuity for UEs moving from one cell to another, security and mechanisms for priority and pre-emption.

Assuming the eMBMS framework and GCSE as basis, 3GPP indicates three different architectural approaches that could be adopted~\cite{23768}. The difference among these depends on the network element that manage the GCs. The first defines that the EPC handles the GCs, while the second approach specifies a decentralized solution where the management of GCs is in charge of the eNB. Finally, the third approach is based on IMS.

\section{A GCs promising Solution for a Public Safety LTE network}

In accordance with the indications of the 3GPP, we outline here a promising solution for GCs in the future LTE-based PMR networks and support our vision by providing suitable performance evaluations and comparisons. 

The envisaged solution is based on eMBMS framework and GCSE as previously described, and assumes a centralized approach for the call management (i.e., performed by the EPC) that leads to benefits in terms of coverage and reliability. 

In particular, we will discuss herein two possible approaches. The first is fully based on multicast transmissions, named \textit{static eMBMS activation}, where the unicast mode is considered only as backup solution when the multicast transmission is not available, e.g., due to bad propagation conditions or because the UE moves out of a MBMS area. Unicast mode is activated only to provide service continuity when the UE detects an increasing packet loss. 

A second, more advanced solution, named \textit{dynamic eMBMS activation}, is considered in order to increase the system flexibility and efficiency. \textit{Dynamic eMBMS activation} manages group communications over both unicast and multicast transmissions. The idea is based on the ability to perform a counting procedure for the GCSE, with the aim to transmit downlink media contents through multiple unicast bearers until a certain number of users requires the same GC service, hence to activate a multicast bearer. The advantage is that the GC is never broadcast in a cell with no active group members, and, hence, the spectrum efficiency is maximized.

\begin{figure}[tbd] 
\begin{center}
\includegraphics[width=\columnwidth]{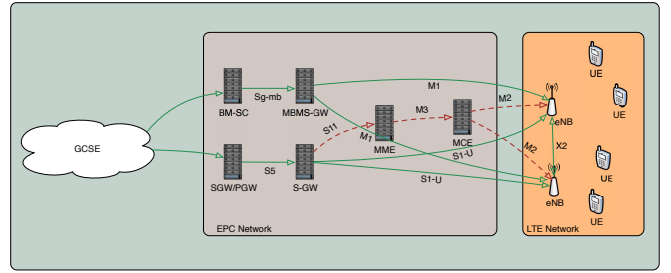}
\caption{System view of PMR network.}
\label{fig.network}
\end{center}
\vspace{-2mm}
\end{figure}

For both solutions, the network architecture is represented in Fig. \ref{fig.network}, where the GCSE acts as an \textit{Application Server} (GCSE AS) providing the functionality for the management of GCs. In addition, the GCSE uses the existing LTE standardized interfaces in order to communicate with the EPC and selects the proper transmission scheme for the delivery of GCs. Finally, multicast bearers are considered for downlink communications whenever possible. On the other hand, uplink traffic is always sent via unicast bearers.
GCSE provides a multicast to unicast switching mechanism depending on the selected approach. Hence, the GCSE is connected to both the PGW (for the distribution of unicast traffic) and to the BM-SC (for the multicast traffic) through the SGi and GC2 interfaces, respectively. 

\subsubsection{Static eMBMS activation}
In the \textit{static eMBMS activation} solution, we assume that the GCs are always managed by resorting to multicast transmission except when it is not available. This approach  exploits the basic structure of the eMBMS, for which it is needed to verify the satisfaction of the PSS requirements.
From Fig. \ref{fig.rate_gc} it is possible to see that this is true for the expected requirements in terms of number of GCs supported for each type of service \cite{lewpMatrix}. 

However, the most critical requirement is represented by the GC set-up time that must be lower than 300 $\milli\second$ (Sec. \ref{REQ}). This depends on the time requested to establish the eMBMS bearer and to start-up the call. Two options have been analysed:

\begin{itemize}
\item  \textit{Pre-established eMBMS Bearer}. The network establishes the eMBMS bearer over preconfigured MBMS areas before the GC starts. This implies that the BM-SC pre-establishes in advance all the information related to the GC, such as the \textit{Temporary Multicast Group Identifier} (TMGI), QoS class and the eNBs belonging to the MBMS area. In particular, the GCSE AS requests the creation of an eMBMS bearer to the BM-SC by means of the PCRF interface, which is in charge of the exchange of the information related to the eMBMS session.  As soon as a UE requests a GC, the downlink traffic is transmitted using one of the pre-established eMBMS bearer. This solution provides a fast total set-up time for the GC service that depends only on the time to start-up the call. 3GPP estimates that it is nearly 220-250 $\milli\second$~\cite{36868}, which is consistent with the requirement for the PMR voice communication (see Sec. \ref{REQ}). 
\item  \textit{Dynamic bearer setup at group call start-up}. In this case the eMBMS bearer is established only when needed. It means that in addition to the 220-250 $\milli\second$ for the call set-up, also the delay for the downlink bearer establishment shall be considered for evaluating the user experience. The additional latency is assessed on the order of 115 $\milli\second$ \cite{36868}, taking into account 10 $\milli\second$ for radio interface delay, 5 $\milli\second$ for network interface delay ad 5 $\milli\second$ of request processing delay. This additional latency is not negligible and the total delay experienced by users, for call set-up and downlink bearer set-up, also exceeds 300 $\milli\second$. 
\end{itemize}

\begin{figure}[tbd]
\begin{center}
\includegraphics[width=\columnwidth]{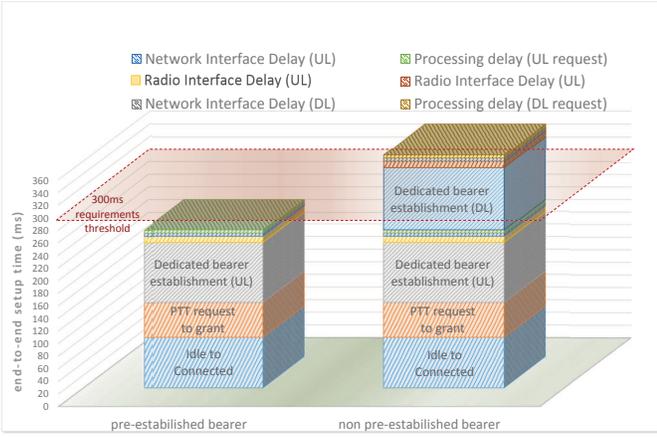}
\caption{End-to-end setup latency for the two considered case with and without pre-established downlink bearer.}
\label{bearer_gr}
\end{center}
\vspace{-2mm}
\end{figure}

Fig. \ref{bearer_gr} shows the delay contributions composing the overall end-to-end latency at call start-up, for both options.

Even if dynamic bearer set-up is more flexible and permits an optimization of the user resources, pre-established bearer is the option selected in our solution, because it permits to satisfy the GC set-up time requirement. 

However, it could be not sufficient because when the MBMS service is mapped on the eMBMS bearer and on the logical MTCH channel, it has to be multiplexed with the control information (MCCH) and then sent on the physical MCH to be transmitted. In the current LTE standard, the MCCH can be updated with a minimum period of 5.12 $\second$ (called \textit{MCCH modification period}), it means that the GC should wait also this time before to be transmitted exceeding the set-up delay requirement.
Hence, we propose also a shorter MCCH modification period that should be about 50 $\milli\second$ (once every 5 LTE frames).

\begin{figure}[tbd]
\begin{center}
\includegraphics[width=\columnwidth]{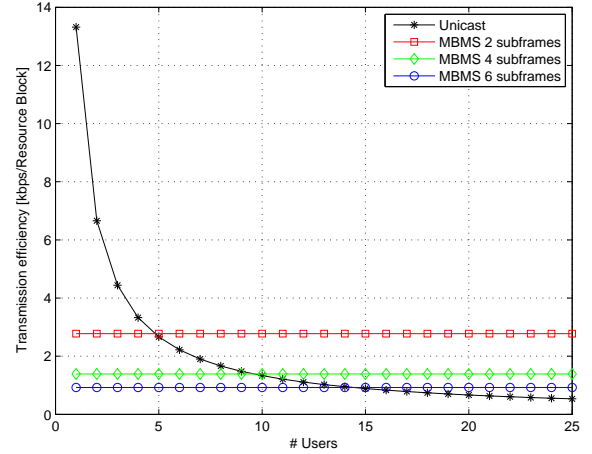}
\caption{Spectral efficiency evaluations for dynamic eMBMS activation.}
\label{tx_eff}
\end{center}
\vspace{-2mm}
\end{figure}

\subsubsection{\textit{Dynamic eMBMS activation}}
In this case the use of unicast and multicast is more flexible and also the GC set-up time requirement can be relaxed. Indeed in this solution UEs are always served through unicast transmissions at the communication start-up, and the use of pre-established unicast bearer shall be considered in order to accomplish the required end-to-end call set-up time. On the other end the activation of eMBMS bearer can be dynamic, since users are already receiving the service and will be able to switch on multicast reception as soon as eMBMS has been completely activated, as a consequence the MCCH modification period can be neglected and unmodified.
The number of users associated to a GC is provided on a per-cell basis by the PGW by means of the \textit{User Location Information} (ULI) procedure. Consequently, the GCSE selects the most efficient transmission scheme. 
An alternative to the ULI procedure is represented by the UE that sends a message to the GCSE whenever moves from one eNB to another one. This solution implies a higher data transmission over the core network, but at the same time allows the UE to switch to IDLE mode in order to save battery life delivering high burst of data in short period of time. 

In the proposed \textit{Dynamic eMBMS activation} solution the GCSE exploits the information regarding the UEs associated to a GC service, to detect the most efficient solutions. In particular, the total amount of resources requested by the UEs if they would be served with unicast mode is evaluated and compared with the resources reserved for the multicast mode. 
As an example Fig. \ref{tx_eff} shows the spectral efficiency in terms of throughput of a group member, normalized to the resources allocated to the multimedia GC service when the number of group(s) members increases. We can see that the multicast solution is independent on the number of group members, the resource usage depends only on the number of subframes allocated to the multicast service. Conversely, the unicast spectral efficiency decreases with the number of served UEs. With the proposed \textit{Dynamic eMBMS activation} procedure the system is able to switch from unicast to multicast when the number of UEs overcomes a given value represented by the intersection of the curves, hence the spectral efficiency is always the maximum value (i.e., the envelope of the unicast and multicast curves). 

\section{Conclusions}\label{CL}
The development of the LTE technology offers an excellent opportunity to improve both performance and capabilities of the actual PMR communication systems. However, towards this goal, substantial research efforts are needed mainly to support the specific critical features of PSS systems currently not available in the LTE standard. After a critical review of the state-of-the-art concerning Group Call standardization proposals in 3GPP, the paper outlined a viable architecture solution and validated its efficiency by providing performance evaluations.  

\bibliographystyle{IEEEtran}

\begin{IEEEbiography}{Lorenzo Carl\`a}
(S'13) received a M.Sc. in telecommunications engineering from the University of Florence in 2013, and is currently a Ph.D. student the Department of Information Engineering, University of Florence. His research interest are mainly focused on radio resource allocation strategy for broadband wireless networks and digital network coding. He is an author of technical papers published in international conferences.
\end{IEEEbiography}

\begin{IEEEbiography}{Romano Fantacci}
(M'84, SM'90, F'05) received the PhD degree in Telecommunications in 1987 from the University of Florence where he works as full professor since 1999.
Romano has been involved in several research projects and is author of more than 300 papers.
He has had an active role in IEEE for several years.
He was Associate Editor for several journals and funder Area Editor for IEEE Transactions on Wireless Communications.
\end{IEEEbiography}

\begin{IEEEbiography}{Francesco Gei}
received the Master's Degree in Telecommunications Engineering in 2008 from the University of Florence, where he currently collaborate as researcher. After initial studies on UWB (Ultra Wide Band) communication systems (2005-2008), Francesco has been involved in industrial research project on SDR (Software Defined Radios) and mesh networks for military and professional purpose (2008-2013). In last years, he focuses his activity on PMR (Professional Mobile Radio) evolution towards 4G and portability of PMR services over LTE (Long Term Evolution) mobile networks.
\end{IEEEbiography}

\begin{IEEEbiography}{Dania Marabissi}
(M'00, SM'13) received the PhD degree in Informatics and Telecommunications Engineering in 2004 from the University of Florence where she works as an assistant professor.
She has been involved in several national and European research projects and is author of technical papers published in international journals and conferences.
She is Principal Investigator of the FIRB project ``HeLD''.
She was Associate Editor for IEEE Transaction on Vehicular Technology.
\end{IEEEbiography}

\begin{IEEEbiography}{Luigia Micciullo}
received her degree in telecommunication engineering from the University of Florence in 2007. In 2007 she joined the Department of Electronics and Telecommunications of the University of Florence as a research assistant. Her research interests
include physical layer for wireless communications, OFDM, intra-systems coexistence, interference management, and multi-tiered networks, broadband professional mobile radio evolution.
\end{IEEEbiography}

\end{document}